
\input phyzzx
\nonstopmode
\sequentialequations
\twelvepoint
\nopubblock
\tolerance=5000
\overfullrule=0pt

\REF\skyrme{T.H.R. Skyrme, Proc. Roy. Soc. London, Ser. A {\bf 247}
(1958) 260.}

\REF\moduli{A. Belavin and A. Polyakov, JETP Lett.
{\bf 22} (1975) 245.}

\REF\wilczekzee{F. Wilczek and A. Zee, Phys. Rev. Lett. {\bf 51}
(1983) 2250.}

\REF\leekane{D.-H. Lee and C.L. Kane, Phys. Rev. Lett. {\bf 64}
(1990) 1313.}

\REF\sondhi{S.L. Sondhi, A. Karlhede, S.A. Kivelson, and E.H. Rezayi,
Phys. Rev. {\bf B 47} (1993) 16419.}

\REF\macdonald{K. Moon, H. Mori, K. Yang, S.M. Girvin,
A.H. MacDonald, L. Zheng, D. Yoshioka, Phys. Rev. {\bf B 51}
(1995) 5138.}

\REF\halperin{B.I. Halperin, Helv. Phys. Acta. {\bf 56} (1983) 75}

\REF\barrett{S.E. Barrett, R. Tycko, L.N. Pfeiffer, and K.W. West,
Phys. Rev. Lett. {\bf 72} (1994) 1368 and
S.E. Barrett,  R. Tycko, G. Dabbagh,  L.N. Pfeiffer, and K.W. West,
preprint 1994.}

\REF\asw{D. Arovas, J. R. Schrieffer, and F. Wilczek, Phys. Rev. Lett.
{\bf 53}, (1984) 722. }

\REF\witten{E. Witten, Nucl. Phys. {\bf B223} (1983) 422.}

\REF\finkelstein{D. Finkelstein and J. Rubinstein, J. Math. Phys.
{\bf 9} (1968) 1762.}

\line{\hfill PUPT 1581, IASSNS-HEP 95/104}
\line{\hfill cond-mat/9512061}
\line{\hfill November 1995}
\titlepage
\title{Quantum Numbers of Textured Hall Effect Quasiparticles}
\vskip.2cm

\author{Chetan Nayak\foot{Research supported in part by a Fannie
and John Hertz Foundation fellowship.~~~
nayak@puhep1.princeton.edu}}
\vskip .2cm
\centerline{{\it Department of Physics }}
\centerline{{\it Joseph Henry Laboratories }}
\centerline{{\it Princeton University }}
\centerline{{\it Princeton, N.J. 08544 }}

\vskip.2cm

\author{Frank Wilczek\foot{Research supported in part by DOE grant
DE-FG02-90ER40542.~~~wilczek@sns.ias.edu}}
\vskip.2cm
\centerline{{\it School of Natural Sciences}}
\centerline{{\it Institute for Advanced Study}}
\centerline{{\it Olden Lane}}
\centerline{{\it Princeton, N.J. 08540}}
\endpage

\abstract{We propose a class of variational wave functions
with slow variation in spin and charge density and simple vortex
structure at infinity, which properly generalize both the Laughlin
quasiparticles and  baby Skyrmions.
We argue that the spin of the corresponding quasiparticle has a
fractional part related in a universal fashion to the properties of the
bulk state, and propose a direct experimental test of this claim.
We show that certain
spin-singlet quantum Hall states can be understood as arising from primary
polarized states by Skyrmion condensation.}

\endpage

\chapter{Introduction}

Almost 40 years ago
Skyrme [\skyrme] introduced a model of nucleons as distributions of pion fields
which has inspired much work, both in its original
context and more generally in
the quantum theory of solitons.
In particular, a realization in quantum ferromagnets was contemplated
early on [\moduli ].
More than 10 years ago Wilczek and Zee
[\wilczekzee] discussed the novel fractional spin
and quantum statistics and that can arise for what
they called ``baby Skyrmions'' in 2+1 dimensions.  These objects
(which we shall here call simply skyrmions) arise in an $SO(3)$ nonlinear
$\sigma$-model, where they are described by field distributions of the
type
$$
\vec n (r, \phi ) ~=~
(\sin \theta (r) \cos \phi,~ \sin \theta (r) \sin \phi,~  \cos \theta (r) ) ~,
\eqn\skyrfield
$$
where $\vec n$ is a unit vector field, and $\theta (r)$ runs from
$-\pi$ at $r=0$ to $0$ at $r\rightarrow \infty$.

Recently there has been a revival of interest in objects of this
kind, inspired by the important realization
that for some quantum Hall states -- including
the classic $\nu=1$ and $\nu=1/3$ cases --
the lowest energy charged quasiparticles
may be skyrmions  [\leekane,\sondhi,\macdonald,\barrett].
There is significant numerical and
experimental support for this circle of ideas.

The recent literature on skyrmions in the quantum Hall complex
takes as its starting point
an effective theory of the state in question
which was initially deduced from a Landau-Ginzburg theory
of the quantum Hall effect [\sondhi] and has since
received some microscopic justification [\macdonald].
Here, by addressing the determination of quantum numbers in
a more direct fashion, we refine and
partially justify the effective theory.
We find that the traditional Laughlin quasihole finds a natural
place as a zero size texture, and that the skyrmion can be
interpreted as a rotationally symmetric texture modified by a flux
insertion that acts on up spins only.
Most important, we find from our microscopic considerations
that a parameter in the effective quantum theory,
the coefficient of the Hopf term, is quantized, with
a value displaced from integer by a universal constant depending on the
bulk state.  This fact is reflected in quantization in the properties of the
skyrmion as material parameters are varied, and specifically to a
non-integral part of its spin which is in principle
observable experimentally.  (Note that spin in the direction of the
magnetic field, here taken as the $z$ direction, is a good quantum
number. In what follows, when we refer to skyrmion spin we mean this
component.)  We also suggest that
the anomalous quantum
properties of the skyrmions -- their fractional
charge, statistics, and spin -- all
come together in a hierarchical construction, by way
of skyrmion condensation, of quantum Hall states
involving spin degrees of freedom.

\chapter{Wave Function for Spin Textures}


We will be considering the possibility that the energy splitting
between up and down spins, though non-zero due to the background
magnetic field, is not so large as to preclude a dynamical role for
both. In fact, as was pointed out by Halperin [\halperin], this is the
case in GaAs, where the $g$-factor is $\sim {1\over {60}}$.
We suppose that in the bulk state the spins are all aligned
pointing up at infinity, with a density corresponding
to an incompressible state.

To understand how it can be that
the presumed smallness of the  Zeeman
energy for flipping an individual spin does
not lead to a vast proliferation of low-energy excitations, we must
recognize the possible significance of exchange energy, which
makes it costly to have rapid changes in the direction of
magnetization.
Thus one can anticipate that locally there is a well-defined
magnetization direction, and that an appropriate class of
wave functions to describe the low-energy excitations
should in some sense
reflect that locally the physics resembles that of an incompressible
single-spin fluid, but that the direction of the magnetization may
slowly vary.  This possibility of such a procedure is implicitly
assumed in the use of a non-linear $\sigma$ model for the low-energy
dynamics.
However, since the Hall states are both highly correlated and rigid --
described by holomorphic functions -- it is not entirely obvious how, or
perhaps even if, such states can be pieced together.

Our first goal is therefore to show  explicitly, in enough detail to
support our later considerations on quasiparticles, how to proceed in the
special case of a centered geometry.  We will generalize a procedure
used in the spin-polarized case.  Taking the center at the origin,
it is appropriate to work in symmetric gauge.  Then a convenient
basis of single-particle wave functions in the lowest Landau level (to
which, for simplicity, we restrict ourselves) takes the form
$f_l (z) = z^l e^{-|z|^2/4{l_0^2}}$, where ${l_0} = (eB)^{-1/2}$
is the magnetic length.
This function represents a ring of charge of thickness ${l_0} \sqrt{2\pi}$ at
distance ${l_0}\sqrt {2(l+1)}$ from the origin.  Note that successive rings
overlap significantly.  Now to represent the possibility of a
non-trivial dependence of magnetization on distance, we may consider
multiplying this spatial wave function by the spinor $s_l =
\cos ({\theta (l)\over 2 }) u + e^{i\phi (l)} \sin ({\theta (l) \over 2} ) v$,
where
$u=(1,0)^{\rm T}, v=(0,1)^{\rm T}$.
That spinor corresponds to the
magnetization vector
$(\sin \theta (l)  \cos \phi (l) ,~ \sin \theta(l)  \sin \phi (l),~
\cos \theta (l) )$.  So far $\theta (l)$ and $\phi (l)$ are
simply
prescribed
functions of $l$, with no explicit space dependence.
Now define the matrix of spinors
$$
M_{kl} ~=~ f_l (z_k ) s_l (u_k, v_k)
\eqn\wavematrix
$$
and the spinor wave function
$$
\Psi_1 (z_k) ~=~ \det M ~.
\eqn\wavefunction
$$
In these expressions, it is to be understood that the indices $k$ and
$l$ run over $0, ... , N-1$, and that the product of spinor factors is
to be understood as a tensor product.

We claim that $\Psi_1$, as constructed in \wavefunction\ , is suitable
to implement
the physical requirements mentioned above.  That is, it keeps the
charge density uniform (at the density appropriate to a single filled
Landau level) while allowing the direction of magnetization to vary in
space, through the dependence of $\theta $ and $\phi $ on $l$.
Furthermore
if these functions depend slowly on $l$, then the exchange energy will
not be unfavorable, since nearby spins will be aligned.
As long as $\theta$ approaches zero for large
$l$, it will match on to the bulk state at infinity; for $\theta$
identically zero $\Psi$ reduces, of course, to
the standard polarized spin
droplet.  It represents,
for general $\theta $ and $\phi$, a class of low-energy localized spin
texture excitations.

Wave functions, $\Psi_m$, for
spin texture excitations at any primary
Laughlin fraction $\nu = 1/m$ may be constructed in the following way.
We define
$${\Psi_m}= {({\rm det} {\tilde M})^m}\,
{e^{-{\sum_j}|z_j|^2/4{l_0^2}}}\eqn\wavefunctionm$$
where ${{\tilde M}_{kl}} = {z_k^l}\,\,{r_l}({u_k}, {v_k})$.
${r_l}$ is defined only by
${r_{l_1}}{r_{l_2}}\ldots {r_{l_m}} = {s_{{l_1}+{l_2}+\ldots+{l_m}}}$
and $s_l$ is the spinor defined above. When the determinant
is expanded and taken to the $m^{\rm th}$ power, each term
will have $m$ ${r_n}({u_k}, {v_k})$'s and can, hence, be rewritten in terms
of the appropriate ${s_l}({u_k}, {v_k})$, so the
wavefunction \wavefunctionm\ is well defined. As in the case of $\nu=1$,
the amplitude for two electrons to approach each other vanishes
in the limit that $\theta$ and $\phi$ are very slowly varying.
It is noteworthy that the wave function here does {\it not\/} arise by
straightforward flux attachment from the $\nu=1$ wave function, but
requires taking a peculiar ``$m^{\rm th}$ root'' of the spinor.
Straightforward flux attachment, {\it i.e}. putting the entire spinor
dependence
in one factor of \wavefunctionm , would not associate a given spin
direction with a definite spatial position.

Thus far, we have only constructed textures in which the spin
direction is a function only of the radial variable -- that is, $l$.
Dependence on the azimuthal angle, consistent with the constraint of
slow variation in the direction but not the magnitude of the
magnetization, and
with appropriate correlations and holomorphy, can be incorporated
as follows.  In the $l~$th partial wave, instead of a constant spinor
$s_l$ multiplying $f_l$, we must allow for dependence
$s_l(\tilde \phi)~ =~ \alpha_l (\tilde \phi) u+ \beta_l (\tilde \phi)v$
on the spatial
angle
$\tilde \phi$.  This can be achieved as follows.  One has
approximately
$$
\eqalign{
\cos \tilde \phi ~ &\sim ~ {1\over 2 }\biggl({z\over R} +
{R\over z }\biggr) \cr
\sin \tilde \phi ~ &\sim ~ {1\over {2i}}\biggl({z\over R} -
{R\over z}\biggr) \cr }
\eqn\anglesfromz
$$
on the ring where $f_l$ is supported, where $R = {l_0}\sqrt{2(l+1)} $.  If
$\alpha_l , \beta_l $  are slowly varying on the scale of a magnetic
length, then their Fourier expansion with respect to $\tilde
\phi$ will essentially terminate after $\sim \sqrt l$ terms.  Thus,
expressing $\alpha_l , \beta_l $ in terms of
$z$ using \anglesfromz\ , we do not meet
overly large powers of $1/z$ -- the $z^l$ factor in $f_l$ prevents any
singularity from occuring in the product.  Using this procedure within
the partial waves, and
the determinantal construction to piece the partial waves together, we
can indeed accommodate the general slowly varying texture with appropriate
correlations and holomorphy properties.

\chapter{Vortices and Skyrmions in Context}

The configurations we have described so far have essentially uniform
charge density.   In constructing localized charged excitations, we
want to be sure not to change the structure of the state at long
distances.
Experience with the spin polarized state leads us to suspect that this
must be done by adiabatically inserting a unit of flux.
In making the generalization to states with non-trivial spin
structure, however, we are faced with a choice: do both spin up and
spin down see the flux, or only spin up?
In the former case we will simply carve a hole in the charge density, just as
in the spin-polarized analogue.  The latter case is much more
interesting.  Our $l$-dependent spinor becomes
$$
s_l ~=~
z \cos ({\theta(l) \over 2}) u
+ e^{i\phi (l)} \sin ({\theta (l) \over 2}) v~.
\eqn\skyrspin
$$
This represents a very special case of the angle-dependent construction
discussed above, and our previous discussion of how one passes from
the
one-particle spinor function to the correlated many-body wave function
applies
{\it mutis mutandis}.  The resulting state
is characterized by a special symmetry, in that
simultaneous real space and spinor space rotation
$$
\eqalign{
z &\rightarrow e^{i\gamma}z \cr
u &\rightarrow e^{-i\gamma/2}u \cr
v &\rightarrow e^{i\gamma/2}v \cr}
\eqn\symmtrans
$$
simply multiplies it by a phase.
If $\theta (l)$ runs smoothly from $-\pi$ at $l=0$ to $0$ at
$l \rightarrow \infty$, and we set $\phi (l) \equiv 0 $,
the spin texture associated with \skyrspin\
is nothing but the classic skyrmion spin texture, as one sees on
comparing the magnetization direction associated to \skyrspin\ to
(1.1) and  recalling that $l \sim r^2 $.  Of course if
$\theta = 0$ identically we have the classic Laughlin quasihole.
Since the exchange energy is typically very large,
we require the spin to be slowly-varying with position.
In this case, an excitation of the form \skyrspin\ must
have $\theta(0)=0$ or $\theta(0)=\pm\pi$.

By comparing the radius of the droplet for $N$ particles with a
centered flux tube, one readily concludes that this configuration contains
a net density deficit corresponding to $1/m$ electron, whether
the flux tube affects both spins or only spin up, and whatever the
detailed form of $\theta (l)$, so long as it approaches zero as
$l \rightarrow \infty$.

Now we are in a position to appreciate, following [\sondhi ],
the possible energetic
advantage of the skyrmion -- or more general -- textures
for accommodating charge
inhomogeneities.
For whereas the classic Laughlin quasihole involves a density
inhomogeneity on the scale of the magnetic length, and thus a heavy
price in repulsive energy, the skyrmion texture allows the
inhomogeneity
to be spread over its physical radius, {\it i.e}. the size of the
region over which $\theta$ differs significantly from zero.  Under
appropriate conditions, this
advantage can be worth the price in unfavorable Zeeman and exchange
energy.
Note that for the quasiparticle (as opposed to the quasihole) it is
natural
to make the choice $s_l = z^{-1} \cos ({\theta(l) \over 2}) u
+ e^{i\phi (l)} \sin ({\theta (l) \over 2} ) v$; as long as
$\cos{ \theta (0) \over 2 }$ vanishes this introduces no singularity
(as in the case of the skyrmion, this avoids rapid variations
in the direction of the spin),
and associates an antiskyrmion spin texture with the quasiparticle.
In the spin-polarized case the quasiparticles cannot be implemented in
quite such a simple way.
Of course, the energetically favored forms for $\theta (l)$  have
every reason to differ microscopically between quasiholes and quasiparticles.
The antiskyrmion is also symmetric under combined real space and
spinor space rotations, but the spinor space rotation is the
opposite to that of the skyrmion.
This symmetry has important consequences, as we shall discuss momentarily.

Let us briefly indicate how these considerations on the microscopic
theory might be incorporated in an effective theory.  We seek to describe
the low energy excitations of the incompressible drop
as spin textures, and in particular to consider how coupling of real
electromagnetic and `fictitious' statistical gauge fields governs the
charge and statistics of the elementary excitations.  Given the spinor
field $\psi_i (z) \equiv (u(z), v(z) )^{\rm T}$, there are two
candidate conserved currents to which a $U(1)$ gauge field could
couple, {\it viz}.
$$
\eqalign{
J^\mu_{\rm skyrme} ~&=~
	\epsilon^{\mu\nu\rho} \partial_\nu {\bar \psi}^i
	\partial_\rho \psi_i~, \cr
{\tilde J}^\mu ~&=~
	\epsilon^{\mu\nu\rho} \epsilon^{ij} \partial_\nu \psi_i
	\partial_\rho \psi_j ~. \cr}
\eqn\currents
$$
The latter symmetry is not a suitable charge current because it
is odd under the spinor space rotation $u\rightarrow -u$.
Hence, the current in an effective field theory can only be given by
$J^\mu_{\rm skyrme}$. In particular, this theory should have a term
$${{\cal L}_{\rm Hopf}} = - {4\pi \over m}\Bigl({J^\alpha_{\rm skyrme}}
{a_\alpha}-{1\over2}
{\epsilon^{\alpha\beta\gamma}}{a_\alpha}{\partial_\beta}
{a_\gamma}\Bigr)\eqn\efflag$$
If the effective Lagrangian is written in terms of the local spin field,
this term is just the Hopf invariant of the corresponding
map $S^3 \rightarrow S^2$. Such a Lagrangian was
discussed in [\sondhi] with gradient energy, Zeeman energy,
and Coulomb repulsion terms. We can redefine
${\tilde a} = {4\pi\over m}\,a$,
so that the Chern-Simons term is
conventionally normalized and the quantized parameter appears explicitly
as a coefficient in the Lagrangian.  Its quantization is connected
with the invariance of the
action under large gauge transformations [\witten ].
Given the mathematical result \efflag,
the ribbon argument of [\finkelstein ,\wilczekzee] applies, and
the anyon character of the skyrmion follows.

\chapter{Quantization of the Spin}

The discussion so far is incomplete in one important respect:  since
a generic particular texture configuration $\Psi (z_k; \alpha_k , \beta_k )$
has no definite value of the spin in the down direction,
the corresponding state is always embedded in a highly degenerate continuum.
One constructs states of definite spin
by forming appropriate
superpositions of these degenerate states, in the manner:
$$
\Psi^s (z_k; \alpha_k , \beta_k ) ~=~
\int^{2\pi}_0 {d\lambda\over 2\pi}
	e^{-is\lambda} \Psi (z_k; \alpha_k, e^{i\lambda}\beta_k  )~.
\eqn\spinstates
$$
The resulting state has exactly $s$ reversed spins.  Now as we have
seen the fundamental charged particles generically feature
a localized fractional
charge,
as is familiar in the fully polarized case (and for the same reason).
Since the total particle density has a fractional piece, and the
number of reversed spins is integral, clearly the net spin relative to
the ground state is fractional.

The fractional part of the spin is intimately related to the anyonic
statistics of the skyrmion as a result of its symmetry \symmtrans\
under simultaneous real space and spin space rotations.
By our previous arguments, skyrmions and anti-skyrmions
have statistics $-{1\over m}$; as a result of the
spin-statistics connection [\wilczekzee], they have intrinsic
angular momentum $-{1\over {2m}}$. On the other hand, the special symmetry
\symmtrans\ of
the skyrmion implies that it is an eigenstate of $L-S$ with
integer eigenvalue, where $L$ is the intrinsic angular momentum
and $S$ is the spin -- which is just an internal
quantum number in this context.
Hence, the fractional part of its spin
is equal to the fractional
part of its intrinsic angular momentum. The anti-skyrmion is also symmetric
under a combined real space and spin space rotation,
but one in which the spin rotation is opposite to
the real space rotation, so its spin is equal and opposite to its angular
momentum.

The question
which $s$ is favored for low-lying charged quasiparticles
in a given material is a non-universal question, whose answer
depends on the
detailed form of the Hamiltonian -- that is, it involves energetics, not
merely topology.  Indeed strictly speaking
one should consider the possibility that the
optimal starting
wave function depends on $s$, similarly to how rotation of a molecule
can affect its shape -- ro-vibrational coupling -- although we expect
such effects to be small.   In any case,
one expects
to find that the $s$ which minimizes the
energy for a quasihole exhibits jumps as one changes the in-plane
${\bf B}$ field or material parameters
such as density, impurity concentration, temperature, or well size in the
third direction.  This effect suggests a method of checking the fractional
quantization of the spin.  Indeed, using nuclear magnetic resonance one
can measure the Knight shift induced by a skyrmion, which is proportional
to its spin [\barrett].
If the favored value of the spin jumps by an integer in
response to a small change in the control parameters, then by taking the
ratio of Knight shifts before and after the change one could infer
the ratio, which is of course sensitive to the fractional displacement.
In a material that is not perfectly homogeneous, one might find stable
skyrmions with different values of $s$ at different positions; and at
finite temperature one expects to find each $s$ value represented with
appropriate statistical weight.

\chapter{Skyrmion Condensation and the Hierarchy Construction}

The exotic spin of the skyrmions allows us to understand
spin-singlet states and, more generally, non-polarized states
as hierarchical states resulting from the condensation
of skyrmionic quasiparticles on a polarized parent state.
To see why this is non-trivial, recall that, in the
hierarchy construction, the state at $\nu=2/5$ forms when
charge  $-e/3$ and statistics $-\pi/3$ quasiparticles
of the polarized $\nu=1/3$ state condense in a Laughlin
state.If these quasiparticles are, in fact,
skyrmionic, then the additional
$\nu={2\over5}-{1\over3}={1\over{15}}$
can cancel the spin of the $\nu=1/3$ parent.
More generally, a daughter state
in which skyrmions of a Laughlin state condense
will have charge and spin filling fraction:
$$\nu\,\,=\,\,{1\over m}\,\, + \,\,{\alpha\over m}\,\,
{{1/m}\over{2p-{\alpha/m}}}
\eqn\cfillfrac$$
$${S_z}\,\, =\,\, {1\over2}\,\times\,{1\over m}\,\, - \,\,
\Bigl(J - {\alpha\over {2m}}\Bigr)\,\times\,
{{1/m}\over{2p-{\alpha/m}}}\eqn\sfillfrac$$
where $\alpha=\pm 1$ according to whether skyrmions or
anti-skyrmions condense. $J-{\alpha\over {2m}}$ is the
spin of the skyrmion or antiskyrmion. Observe that the fractional
part of the spin is either aligned or anti-aligned with the
parent, depending on whether it is particle- or hole-like ($\alpha=\pm 1$),
but the integer part is always anti-aligned because it
involve flipping spins of the parent condensate.
This state will have ${S_z}=0$ if $J=p$.\foot{Lee and Kane [\leekane]
also suggested that spin-singlet states could arise
from skyrmion condensation, but the states that they construct are
at denominators such as $1/2$ and $1/4$ whereas our states
are at the same fractions as those of the usual hierarchy,
such as $2/3$, $2/5$, etc.}
It is natural that
the most favorable skyrmion size, $J$, be determined by the
skyrmion inverse density, $p$, in the low Zeeman energy,
high-density limit, where inter-skyrmion interactions
are the limiting factor. This picture for the ${S_z}=0$
states at $\nu={{2p}\over{2pm\pm 1}}$
motivates trial wavefunctions for these
states which are completely analogous to those
of the polarized hierarchy but with skyrmions
substituted for the Laughlin quasiparticles.

Note: This paper supersedes ``Quantum Numbers of Hall Effect
Skyrmions'',
distributed as PUPT 1540, IASSNS 95/35, and cond-mat/9505081 .
We wish to thank an
anonymous referee for pointing out the inadequacy of
our earlier construction.

\endpage

\refout

\end